\documentclass[preprint,12pt]{elsarticle}
\usepackage{graphicx}
\usepackage{color}
\usepackage[colorlinks,bookmarks=false,citecolor=blue,linkcolor=red,urlcolor=blue]{hyperref}
\usepackage{amssymb,amsbsy}
\usepackage{amsmath}
\usepackage{hyperref}   
\usepackage{subfig}
\usepackage{float}
\newcommand{\xref}[1]{\protect\ref{#1}}

\renewcommand{\eqref}[1]{Eq.~(\protect\ref{#1})}
\newcommand{\figref}[1]{Fig.~\protect\ref{#1}}

\newcommand{\rev}[1]{#1}

\journal{Physica E}

\begin{document}

\begin{frontmatter}

\title{Comparison of various schemes to determine the Young's modulus of 
disordered carbon nanomembranes compared to crystalline graphene}
\author{Levin Mihlan\fnref{bi}}
\author{Julian Ehrens\fnref{bi}}
\author{J{\"u}rgen Schnack\corref{cor1}\fnref{bi}}
\ead{jschnack@uni-bielefeld.de}
\cortext[cor1]{corresponding author}
\address[bi]{Dept. of Physics, Bielefeld University, P.O. box
  100131, D-33501 Bielefeld, Germany}

\begin{abstract}
The determination of mechanical properties such as the Young's modulus
provides an important means to compare classical molecular dynamics simulations
with materials. In this respect, ultra-thin materials hold several challenges:
their volume is ambiguous, and different methods to determine a stress-strain 
relation deliver different result in particular for disordered systems. 
Using the example of carbon nanomembranes we \rev{discuss three common approaches
to the problem and} show that 
stress-strain simulations following experimental setups 
deliver correct results if adjusted carefully.
\rev{We provide step-by-step instructions how to perform
trustworthy simulations.}
\end{abstract}


\end{frontmatter}

\section{Introduction}
\label{sec-1}

Mechanical properties such as the Young's modulus (tensile modulus) constitute important observables 
for membranes and other quasi two-di\-men\-sional (2d) materials. The Young's modulus describes 
how a material reacts to certain strain. Although this seems to be a rather global
property it may depend strongly on details of the interaction matrix between atoms of
the material. This is in particular true if the interactions are very different as 
for instance in disordered materials. 

On the theoretical side such materials are often modelled by means of classical 
molecular dynamics schemes since the number of atoms needed for a trustworthy simulation
is way too big for a quantum calculation for instance by means of density functional
theory (DFT). Again this holds in particular for disordered, i.e., non-crystalline systems such as 
bio-molecular membranes or carbon nanomembranes (CNMs)
\rev{in contrast to regular structures such as boron nitride nanosheets on top of graphene domains 
that are crystalline and investigated by means of DFT in \cite{LOA:RSCA:21,OLF:RSCA21}.
In the present paper, disordered  carbon nanomembranes will be treated,
see \figref{CNM} below.} 
As a side remark, mechanical properties are valuable observables for
such systems since due to the lack of a quantum mechanical treatment observables 
such a conductivity or band structure are not available.
\rev{However, the technical problems addressed in this paper apply to both
regular as well as irregular quasi two-dimensional systems.}

In the present article, we follow investigations as e.g.\ outlined in \cite{CDB:MS17,GEH:PE18}
and discuss the evaluation of the Young's modulus of carbon nanomembranes with general
lessons for other 2d materials. Carbon nanomembranes are stable 
quasi two-dimensional disordered 
carbon membranes that are synthesized from aromatic or aliphatic precursor molecules 
grafted on 
gold surfaces \cite{GSE:APL99,LEK:CR05,TBN:AM09,AVW:ASCN13,TuG:AM16,Tur:AdP17,DWN:PCCP19,WEG:2DM19,SWA:JPCC19,EGV:PRB21,SKY:BN22}. 
Their mechanical properties are determined by means of bulge 
experiments or via nano-indentation \cite{ZBG:B11}. 
Amorphous carbon is investigated along similar lines \cite{LXS:ASS13}.

CNMs are rather soft compared to graphene; CNMs have got a modulus 
of about $10$~GPa \cite{ZBG:B11}, whereas graphene features $1000$~GPa 
\cite{FPJ:PRB07,LWK:S08,GEH:PE18}. The structure of
CNMs is irregular and contains many holes through which for instance 
water permeates \cite{YDB:ACSnano18,YHQ:AM20,ASJ:CMS21}. Since CNMs constitute metastable excited 
states (graphite would constitute the ground state) a trustworthy 
simulation of mechanical properties is challenging compared to crystalline 
structures such as graphene or diamond.

In this paper, we discuss the determination of the Young's modulus using three
different \rev{common} procedures. For crystalline samples these methods 
yield very close results \cite{GEH:PE18}. 
However, for disordered materials featuring both strong and weak bonds between atoms, 
where smallest deformations can push a system into new configurations \cite{PSH:CMS24}, 
the three methods deliver rather different outcomes. 
Our conclusion is that both stress-strain based methods discussed in this paper
are suited to calculate the elastic properties of crystalline and amorphous structures.
\rev{For practitioner, we provide step-by-step instructions how to perform
trustworthy simulations.}

The article is structured as follows. Section \xref{sec-2} introduces methods and 
challenges, section \xref{sec-3} provides instructions on how to perform the methods 
and discusses the results obtained with the test structures. 
Section \xref{summary} finishes off the article with a summary.

\section{Methods and Challenges}
\label{sec-2}

\subsection{Molecular Dynamics Simulations}

For our calculations we use the LAMMPS package \cite{TAB:CPC22}. In order to model carbon systems 
the Environment-Dependent Interataomic Potential (EDIP) of Marks is employed \cite{Mar:PRB00}. 
This potential outperforms many of the historic carbon 
potentials, compare e.g.\ \cite{GEH:PE18,TAJ:C19}. 

In general, the application of classical molecular dynamics works for problems discussed in this paper. 
However, one should keep in mind that there are limitations in the application of classical molecular dynamics
for very low temperatures for several reasons. Firstly, it it possible that the system equilibrates 
into metastable states, as there is not enough kinetic energy to overcome potential barriers. 
Thus, the system ends up in a state of non-minimal energy \cite{KaP:N90}. 
Secondly, quantum effects could also play a role at temperatures close to absolute zero \cite{FrS:2001}. 
These phenomena could affect calculated properties at very low temperatures.

In the following, LAMMPS syntax is presented in \textit{italic}.

\begin{figure}[ht!]
\centering
\includegraphics[width=0.6\textwidth]{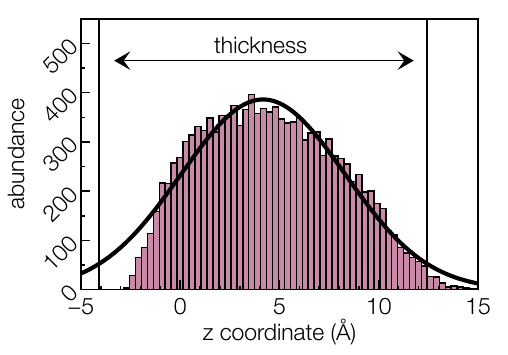}
\caption{An example of a thickness calculation by determining 
the density profile of the structure along $z$-direction perpendicular 
to the membrane. \label{zdensity}}
\end{figure}
 
\subsection{Volume ambiguity of quasi 2d materials}
\label{sec-2-1}

To calculate elastic properties, such as the Young's modulus, 
using a molecular dynamics simulation approach, the stress or pressure 
of the system is usually required, which in turn depends on the volume. 
Quasi two-dimensional (2d) structures, as discussed here, 
may have thickness fluctuations of the same order of magnitude as the thickness itself, 
which complicates the volume prediction \cite{EGV:PRB21,Mihlan:B21}.
Since the LAMMPS modifier \textit{compute stress} always uses the box volume, 
results will be inaccurate and a correction is necessary.

There are two option to address this issue: (a) The visualization program OVITO 
offers a modifier named `construct surface mesh'; it depends on two input parameters 
(radius, smoothness) and delives the volume as output \cite{Ovito:UM}. 
The correction factor then is the quotient of box volume and surface mesh volume.
(b) A second way to estimate the actual volume of the membrane
is to calculate the density profile of the structure in $z$-direction 
and then to use the width of the distribution as thickness. Based on the chosen threshold, 
a fitting $\sigma$-rule (FWHM or $2\sigma$ e.g.) has to be specified, compare \figref{zdensity}.
Again, the correction factor 
is obtained by calculating the quotient with the thickness of the simulation box. 

Regardless of which method is used to determine the volume or thickness, 
both values suffer from a high degree of uncertainty. 
Therefore, a comparison with experimental values should always be approached with caution. 
Nevertheless, a viable trend in the data will be visible.

\subsection{Scaling approach}
\label{sec-2-2}

The Young's modulus $E$ at zero temperature can be evaluated from the curvature 
of the potential energy $U$ at the respective configuration. 
Kinetic effects do not play a role here \cite{HGB:PRL98,ITO:JPCS14}. 
The modulus is obtained as
\begin{equation}
\label{E-2-2}
E_\alpha = \frac{1}{V_0} \left(\frac{\partial^2 U}{\partial \alpha^2}\right)_{\alpha=1}
\ ,
\end{equation}
where $\alpha$ is the factor by which all
positions are scaled along the direction of the dimensionless
unit vector $\vec{e}_{\alpha}$, 
i.e.\
\begin{equation}
\label{E-2-2b}
\vec{x}_i \rightarrow \vec{x}_i + (\alpha - 1) \
\vec{e}_{\alpha}\cdot\vec{x}_i \ \vec{e}_{\alpha}
\ .
\end{equation}
$V_0$ denotes the volume of the sample in equilibrium.

\subsection{Stress-strain method}
\label{sec-2-3}
The stress-strain variant of determining the Young's modulus (tensile modulus) is inspired 
by the similar macroscopic tensile experiments in material science, in which the material 
to be tested is clamped on opposite sides and stretched by a factor $\varepsilon$  at a certain strain rate. 
The stress behavior $\sigma$ depends on the properties of the material.
For the linear regime, the relationship is given by the Young's modulus. 
Here stresses are given by a law similar to Hooke's law $\sigma = E \cdot \varepsilon$ 
where $\varepsilon=(L-L_0)/L_0$ is the strain with $L$ being
the current length and $L_0$ the initial length. Plotting this data, 
the Young's modulus can be determined by fitting a linear function and calculating the slope.

In order to imitate this experiment in an MD simulation, the clamping and the 
stretching of the material must be modeled,
\rev{compare also \cite{Sad:MNL16,SaK:APA16}}. A static approach was chosen 
for this method, meaning no time integration is involved. The procedure can be used 
with periodic or non-periodic lateral boundary conditions, depending on the structure. 
In this case, \textit{p p f} boundary conditions are used, i.e.\ the simulation box 
is only non-periodic in the z-direction. The method presented in Ref.~\cite{Tschopp} 
has been adapted for this procedure by making use of selection box regions 
of the size of the clamp (\textit{region} command in LAMMPS). 
To strain the structure these regions have to move apart. Atoms in the clamp regions 
are excluded from the calculation of stresses by setting the forces to zero 
in their corresponding groups (\textit{fix setforce 0 0 0}) as they are considered as rigid. 
To minimize the error, the clamp size is chosen to be minimal, 
otherwise a systematic error will occur \cite{Mihlan:B21}. 
The clamps are particularly important for non-periodic structures 
as a box deformation does not automatically induce an elongation of the structure.
The initial length of the simulation box along strain direction is saved for 
the calculation of the strain at each time step. Finally, the clamps are moved 
outwards in discrete deformation steps using the \textit{change\_box} command 
with a predefined strain-rate. To ensure that the atoms in the clamped regions also 
follow the movement, the "\textit{remap}" keyword has to be specified in 
the \textit{change\_box} command.

After each deformation step, the potential energy of the structure is minimized 
via the \textit{minimize} command. If laterally periodic boundary conditions are used 
and the Poisson effect is taken care of, 
the structure is given the possibility to relax in the 
non-stretched directions by the command \textit{fix box/relax} 
in order to keep the stress in this direction close to zero, see Fig. \ref{SLG_relax}. 
This command allows to rescale the simulation box in specified directions. 
For this method, the \textit{fix box/relax} is only necessary if the periodicity 
of the unit cell is directly intended by the structure, which is clearly not the 
case for the z-dimension. After the minimization is completed, stress and strain 
are logged for later processing and the next deformation steps are executed 
analogously. The strain rate has to be chosen such that the dynamics 
stays physical, i.e.\ atoms should not move further apart from each other than 
what is covered by the effective potential.

Since no time integration is involved in this calculation, and the structure 
is always minimized with respect to the potential energy, 
it is not a real dynamics simulation but rather a ground state (zero temperature) 
calculation of the Young's modulus. 
The advantage is that numerical errors due to time integration cannot occur.

\subsection{Barostated dynamics}
\label{sec-2-4}

Another and more versatile approach to determine the Young's modulus
by means of molecular dynamics is the simultaneous application of 
simulation box deformation and barostating as suggested in 
Refs.~\cite{CDB:MS17, nylon}. Unlike in the clamp method 
real dynamics is involved here. The Young's modulus 
can then be derived analogously to the stress-strain method discussed 
in Sec.~\ref{sec-2-3}. The difference to the latter method 
lies in how and according to which ensemble the membrane is deformed.

Instead of clamping opposing sides to strain along the, e.g., x-axis of the membrane, 
the length of the simulation box in the respective direction is enlarged 
at a specified strain rate 
$\varepsilon$ without remapping of the atoms, i.e.\ the simulation box is 
enlarged in the chosen direction with no influence on the atom positions, 
see \textit{fix deform} command~\cite{LAMMPS:DOC}, but since it is 
a coherent structure due to the PBCs, each box deformation also 
induces movements of atoms and therefore stress in the material. 
Simultaneously the system is initialized in an isothermal-isobaric ensemble which is 
of Nos{\'{e}}-Hoover type using the 
\textit{fix npt} command \cite{LAMMPS:DOC}, which ensures that the deformation along a specific 
direction results in stresses in only this direction, and the system does not attempt 
to relax via stresses in other spatial directions or via temperature. 
As pressure and temperature are adjustable, the method is able to produce results under different conditions. Examples of stress-strain curves at different temperatures will be presented in 
\figref{stressstrainbaro} in Sec.~\ref{sec-3-3}. 
Here again the Poisson effect is taken care of.

In order for a barostat to control the pressure in a certain direction, 
the simulation must have periodic boundary conditions in this direction, 
which can lead to problems for the z-direction in quasi 2D structures.
All data of the simulated graphene or CNMs have been 
constructed or simulated with PBC in x- and y-directions as standard setup. 
Several approaches for the z-direction have been tested.

\subsection{Investigated structures: graphene and carbon nanomembranes}
\label{structures}

\begin{figure}[ht!]
\centering
\includegraphics[width=0.3\textwidth]{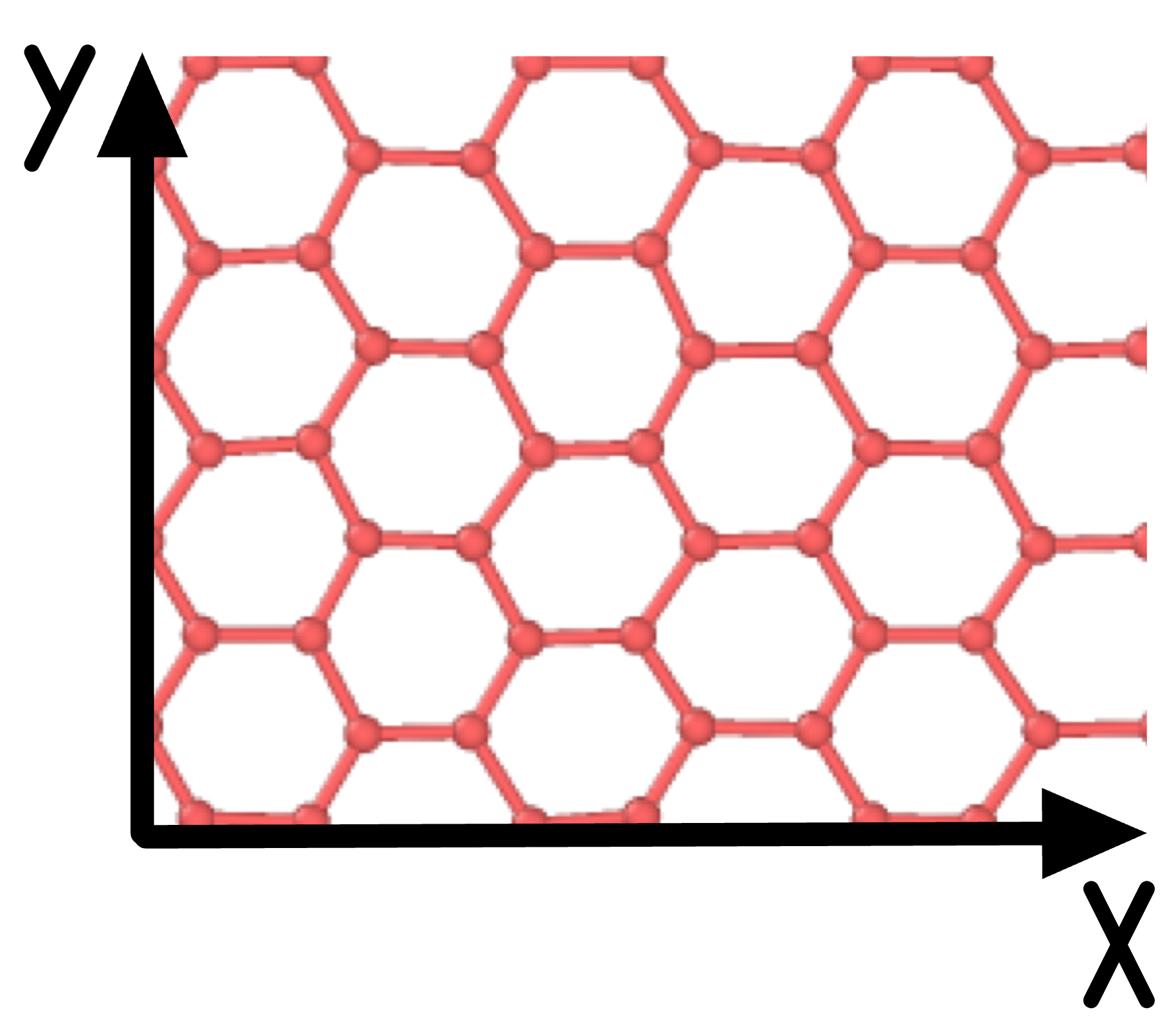}
\includegraphics[width=0.5\textwidth]{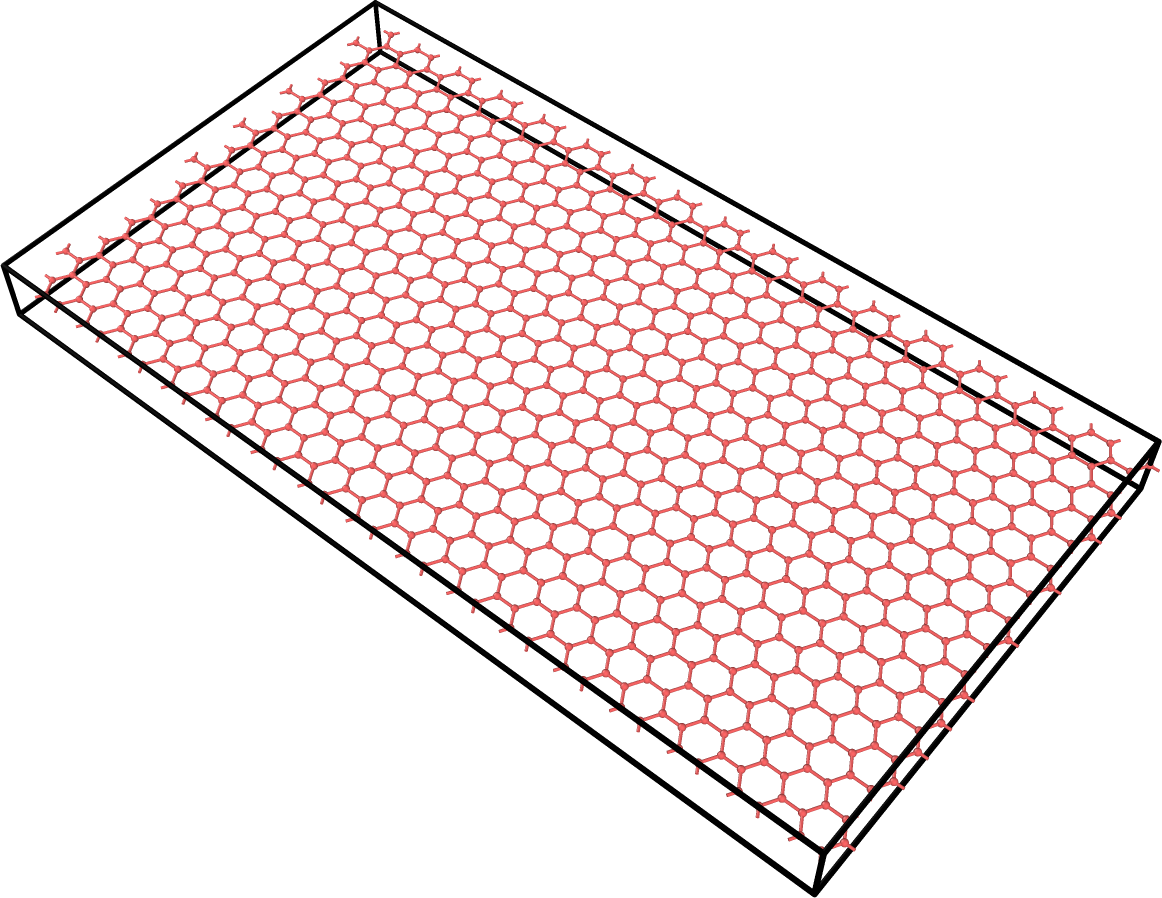}
\caption{Visualisation of graphene data with OVITO \cite{ovito}. 
The x-axis is also called armchair-axis, the y-axis zigzag-axis. \label{graphene}}
\end{figure}

The data set used for graphene generates a xy-periodic box with a size 
of $68.2105$\,\AA\,  $\times$ $36.9306$\,\AA\,  with 960 carbon atoms and an 
inter-atomic distance of $1.421$\,\AA. 
In graphene oriented research the x-direction is often referred to as the armchair- 
and the y-direction as the zigzag-axis \cite{STC:LDSS13}.
A visualisation with OVITO \cite{ovito} is shown in \figref{graphene}. 
Before the data set is used, a simple MD simulation is performed to minimize 
the potential energy of the structure and get rid of possible stresses (\textit{minimize} command). 
Since graphene is a two-dimensional structure, an artificial thickness must be 
chosen for the calculation of the Young's modulus; this is taken 
as $3.35$\,\AA\
as in \cite{GEH:PE18}.

Data for carbon nanomembranes (CNM) are generated using the method discussed in \cite{MrS:BN14,Gayk:M18,Mihlan:B21,EGV:PRB21,Ehrens:D22,Mihlan:M24}.
Again, periodic boundary conditions are used in x- and y-directions. 
The simulation box has a size of $135.147$\,\AA\,  $\times$ $133.801$\,\AA\, and contains $~12500$ atoms. 

\begin{figure}[ht!]
\centering
\includegraphics[width=0.49\textwidth]{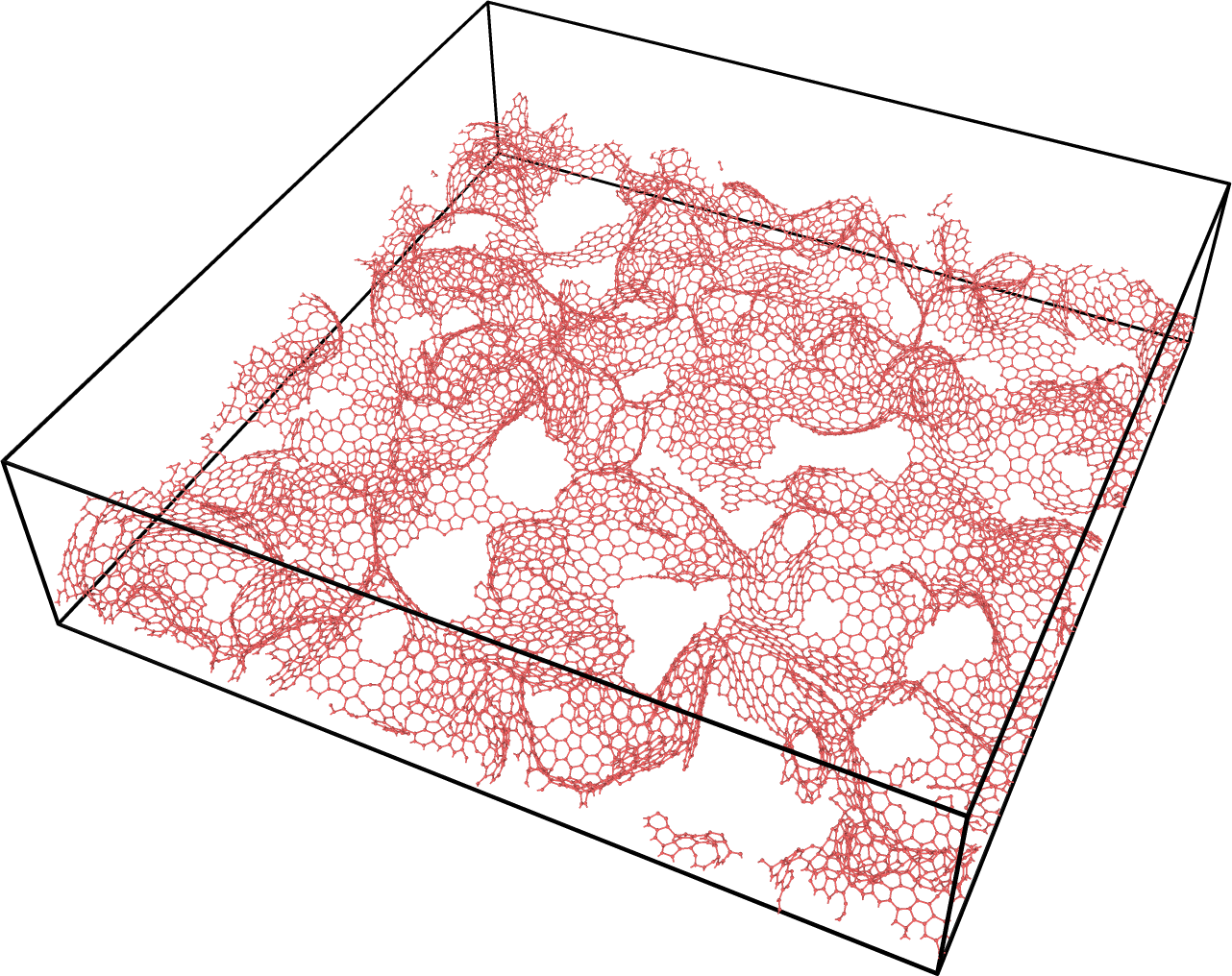}
\includegraphics[width=0.49\textwidth]{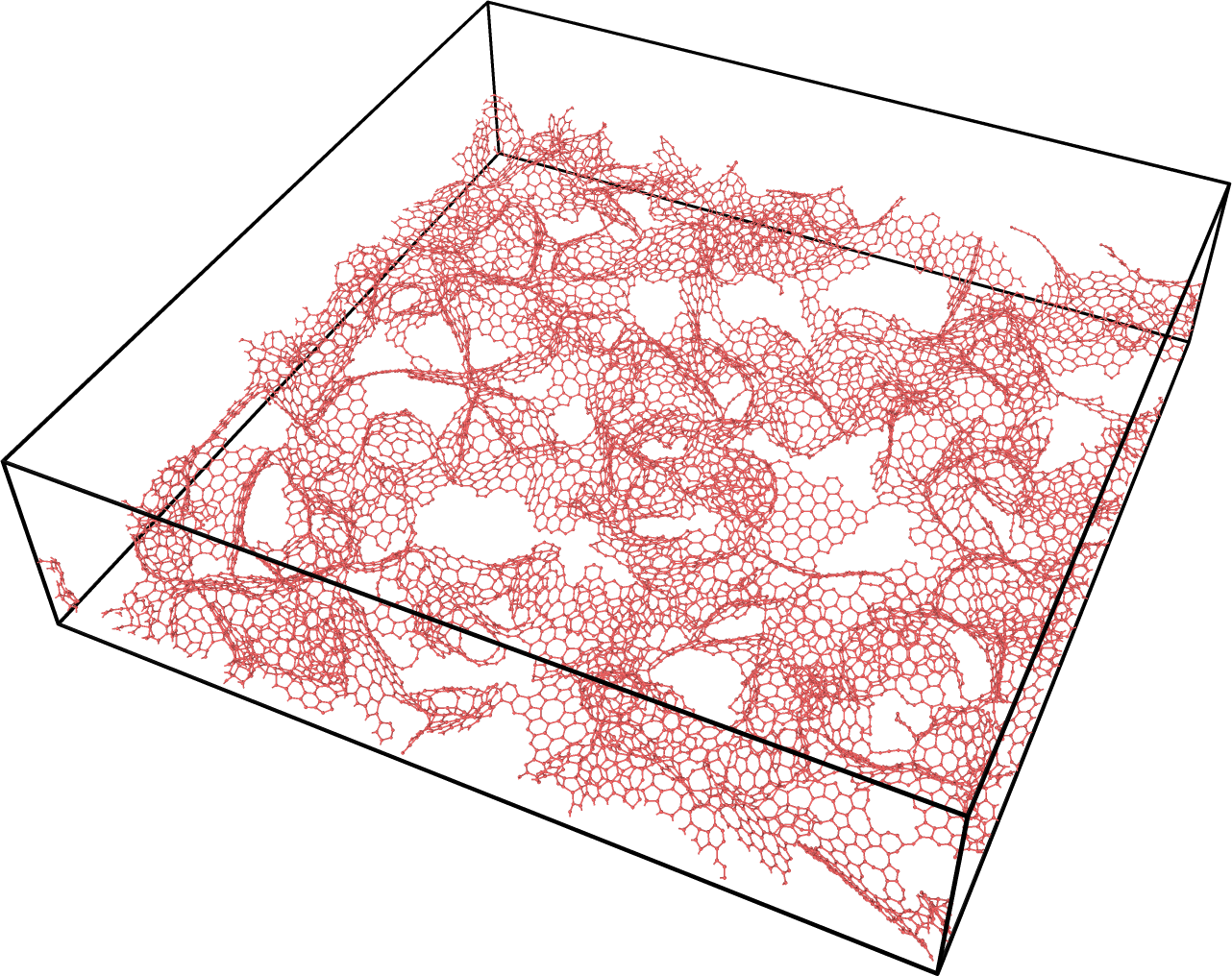}
\caption{Visualisation of TPT-CNM data with OVITO \cite{ovito}. 
Here the simulation box has a height of 33~\AA\, which does not correspond 
to the actual thickness of the membranes. On the left side the membrane is shown 
which will be referenced later with the name CNM1, 
the membrane on the right with the name CNM2.\label{CNM}}
\end{figure}

Because of their amorphous, i.e., irregular structure, 
it is necessary to consider larger model realisations of 
CNMs than would be needed for periodic lattices or crystals.
The size should be large enough to capture all relevant features of the sample.
The thickness of these structures is calculated with the z-density profile 
and a $2\sigma$ fitting rule, as described in Sec.~\ref{sec-2-1}. 
For the structure labeled as CNM1 the thickness is approx. $16.54$~\AA, 
for the structure referred as CNM2 it is $16.07$~\AA. 
It's important to note that the determined thickness is merely an approximation 
and is associated with significant uncertainty. 
However, it is suitable for comparing MD simulations with each other.

\section{Comparison of applied methods}
\label{sec-3}

Here we discuss how to setup simulations and show the results for our investigated structures.

\subsection{Scaling approach}

The scaling approach is the easiest to implement and involves the least computational effort. 
Unfortunately, it provides incorrect values for non-crystalline systems. 
For the graphene test structure this method yields $E_{\text{SLG}}=1088 \pm 11$\,GPa, 
which is in agreement with \cite{GEH:PE18}. 
For the CNMs, however, one obtains $E_{\text{CNM1}}=(240 \pm 5)$~GPa and 
$E_{\text{CNM2}}=(253 \pm 3)$~GPa, 
much larger than experimentally determined.
The scaling approach does not distinguish between strong and weak bonds, 
nor does it recognize other elastic deformation mechanisms, resulting in an overestimated 
Young's modulus in most cases. We therefore consider this method to be unsuitable 
for amorphous media; similar findings were also made by Pashartis et al. \cite{PSH:CMS24}.

\subsection{Stress-strain method}
\label{sec-3-2} 

This method can also be applied to non-periodic structures. 
For non-periodic structures, the clamp size should be selected with care. 
It should be sufficiently large and as small as possible.
Too big clamp sizes result in a predicted mechanical stability that is higher 
than the correct value \cite{Mihlan:B21}. It is advisable to perform
several calculations with different clamp sizes in order to eliminate 
finite size errors by means of an extrapolation to zero clamp size \cite{Mihlan:B21}. 
For structures periodic in the direction of straining the clamp size 
can in fact be chosen to zero, because the box deformation alone generates stresses here.

\begin{figure}[ht!]
\centering
\includegraphics[width=0.49\textwidth]{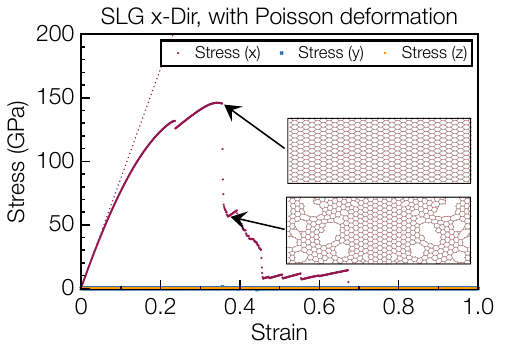}
\includegraphics[width=0.49\textwidth]{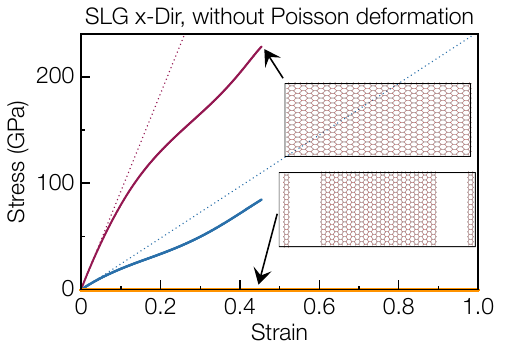}
\caption{Stress-strain diagrams produced with the stress-strain method from section \ref{sec-2-3} 
for single-layer graphene (SLG) in x-direction, the simulation on the left allowed 
for relaxation in y-direction, while the simulation on the right does not, 
which is refered to as uniaxial. This results in Young's moduli of $E_x=861$\,GPa 
for the calculation with relaxation and $E_x=919$\,GPa for the uniaxial calculation. 
For visualisation purposes the data shown here corresponds to a total strain 
of $\varepsilon = 0.8$ and 800 deformation steps. 
The dotted lines correspond to fits of the linear region, 
ranging up to a strain of $\varepsilon = 0.02$ \label{SLG_relax}}
\end{figure}

In most cases, and also here, periodic boundary conditions are used in molecular dynamics. 
Here it is important to allow the non-stretched space dimension to relax its stresses 
if a realistic behavior respecting the Poisson effect is desired. 
The Poisson effect refers to the phenomenon where a material's crosswise strain, 
perpendicular to the direction of applied stress, occurs when it is subjected 
to longitudinal stress, resulting in a reduction of cross-sectional area \cite{Demt:E17}.
This procedure should be repeated in every deformation step 
via the \textit{fix box/relax}-command, described in Sec.~\ref{sec-2-3}. 
This command allows the box to change in size in order to relax. 
If this relaxation is not allowed, greater tension is generated 
in the stretched direction, which results in a uniaxial Young's modulus, see \figref{SLG_relax}. 
The graph shows the different behavior of the graphene structure in response to strain, 
depending on whether relaxation in the non-strained direction is allowed or not. 
For an allowed relaxation, it can be seen that all stress components vanish except 
for the one in the stretched direction, the opposite behavior can be seen if 
all box dimensions remain constant except for the stretched one. 
In addition to the stress component in the x-direction, a stress in the y-direction arises, 
which in turn leads to a higher stress in x-direction, resulting in a higher Young's modulus.

Additionally, the method that allows relaxation exhibits more physically accurate inelastic behavior, 
characterized by numerous lattice rearrangements, etc., while the rigid simulation shows an abrupt 
failure of the structure, see \figref{SLG_relax}. The same behavior is also observed 
for the quasi-two-dimensional CNMs. 
Thus, it becomes clear that for structures periodic in the 
xy-direction, only the case where the unstretched spatial 
dimension is allowed to relax, yields results 
compatible with a realistic tensile test. 

\begin{figure}[ht!]
\centering
\includegraphics[width=0.49\textwidth]{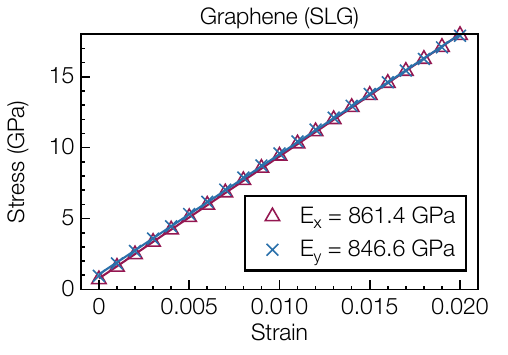}
\includegraphics[width=0.49\textwidth]{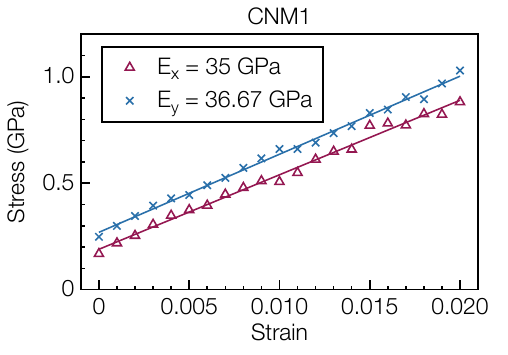}

\includegraphics[width=0.49\textwidth]{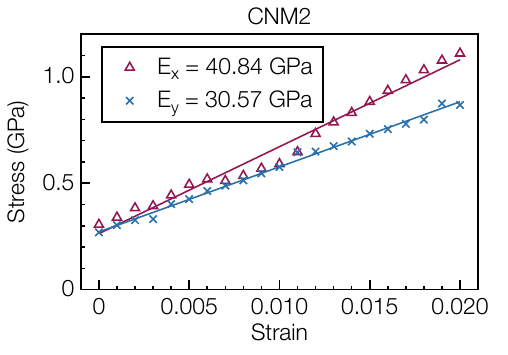}
\includegraphics[width=0.49\textwidth]{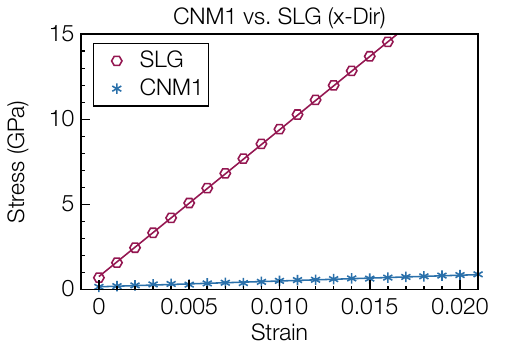}
\caption{Stress-strain diagrams in the elastic regime from structures specified in section \ref{structures}. 
The data points and a linear fit functions are shown for each structure and direction of strain. 
The slope of the linear function corresponds to the Young's modulus 
in GPa and is also noted in the data legend. 
Visualisation date is created with OVITO \cite{ovito}. \label{Ziehen_data}}
\end{figure}

In the following, the results for the structures discussed in Sec.~\ref{structures} are presented. 
Here we use a total strain of $\varepsilon=0.02$ with 20 deformation steps, which lies in the linear range. 
In \figref{Ziehen_data} the stress-strain diagrams with fitted linear functions for the elastic regime 
for both x- and y-directions are provided for single-layer graphene as well as CNM1 and CNM2.
The calculated stresses shown are already corrected with the proper thicknesses. 
All calculations were carried out in a much 
larger simulation box in order to avoid possible errors. 

An isotropic stress behavior is clearly visible for graphene, 
the mean value and its error can be combined 
to 
$E=(854\pm 5.2)$~GPa, which is in general agreement with other literature values \cite{MFD:SuM15}. 
For the carbon nanomembranes the analysis yields 
$E=(35.835\pm 0.835)$~GPa for CNM1 and 
$E=(35.705\pm 5.135)$~GPa for CNM2. As CNMs are amorphous carbon system 
one would expect an isotropic behavior as a stress response. This behavior is 
clearly visible for the CNM1-structure where the two fit functions have
almost identical slopes. The offset between the two data series
\rev{and the fact that the stress is not zero at zero strain} indicates a 
structure that is not fully relaxed in spatial dimensions. 
\rev{In the present case, the offset is ignorable as both materials 
remain well within their linear regime, see \figref{SLG_relax} and \figref{stressstrainbaro}}. 
\rev{However, this is not generally the case, which is why care should always 
be taken to ensure that the material is (practically) fully relaxed before the start of the straining.} 
For the CNM2-structure, two different slopes are evident 
indicating anisotropy in the stress behavior. 
The unit cell may be too small to ensure complete isotropy. 
Therefore, a sufficiently large structure that captures all the features 
of the mother material should always be used.

\subsection{Barostated dynamics}
\label{sec-3-3}

Dynamics at constant pressure and temperature is already widely used in molecular dynamics applications
in order to evaluate the Young modulus \cite{CDB:MS17}. 
However, for 2d materials such as graphene and quasi 2d materials like amorphous carbon 
as well as CNMs 
the application is not trivial. In order to gain deeper insight
various simulations of CNMs and graphene were conducted. Here, the focus is on pressure 
control in the z-direction -- if applicable -- and on the impact this has for
very thin materials.

\begin{figure}[ht!]
\centering
\includegraphics[width=0.49\textwidth]{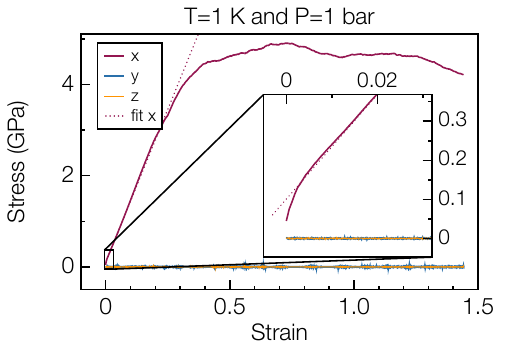}
\includegraphics[width=0.49\textwidth]{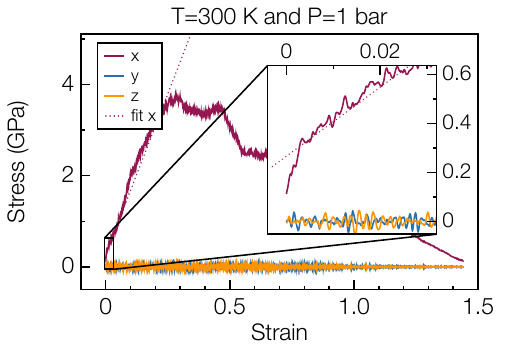}
\caption{Typical stress-strain plots for CNMs at different temperatures 
with the barostat method with a pressure of 1\,bar in unstrained directions. 
Significant fluctuations can be detected at higher temperatures, 
yet unstretched stress components remain around zero. \label{stressstrainbaro}}
\end{figure}

To ensure that the method implemented with LAMMPS delivers correct results 
the Young's modulus of graphene is calculated using the EDIP potential. 
A general practical advise (for every structure) 
is to have the structure equilibrated in the \textit{npt} ensemble 
at desired pressure and temperature before the actual tensile experiment starts. 
This guarantees that there are no initial stresses or temperature differences across the sample. 
\rev{To equilibrate, we use a thermostat and an uncoupled barostat 
in all but the strained direction (in calculations with fixed z this dimension is not included), 
at temperature $T$ and pressure $P=1$\,bar, with damping rates $\gamma_T = 100*dt=0.01$\,ps 
and $\gamma_P=10^3*dt=0.1$\,ps for $15\cdot 10^4$ time steps corresponding to $150$\,ps.}
It is also important to monitor temperature and pressure in unstrained directions 
during the test to ensure that 
\rev{they remain} 
constant within the elastic range, 
see \figref{stressstrainbaro}. 
\rev{This allows us to determine whether the chosen strain rate 
is sufficiently small to accurately represent the tensile test.} 

\begin{figure}[ht!]
\centering
\includegraphics[width=0.49\textwidth]{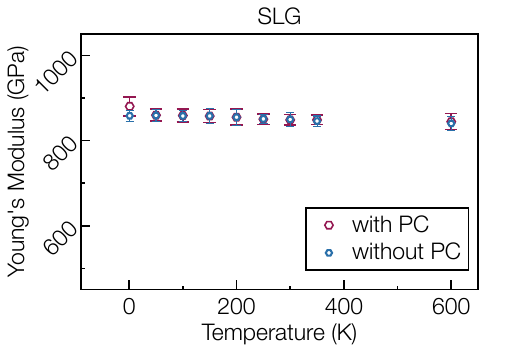}
\includegraphics[width=0.49\textwidth]{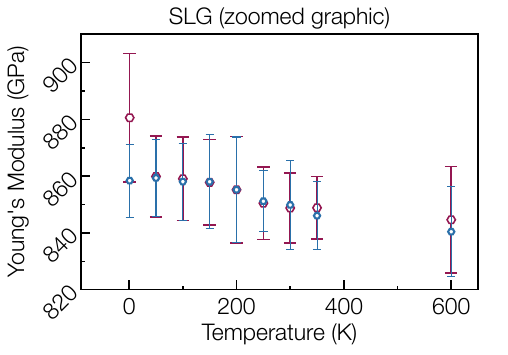}
\caption{Left: Young's modulus depending on temperature, red hexagons correspond to a simulation with 
PBC and pressure control set to 1 bar in z-direction. 
The blue hexagons represent data from simulation with fixed boundary 
and no pressure control in z-direction. Error bars are shown in the corresponding color.
Right: Zoomed in to see smaller deviations.\label{graphene_baro}}
\end{figure}

The thickness of graphene, which is input for the calculation of the modulus, 
is chosen to be $3.35$\,\AA, compare Sec.~\ref{structures}. 
To avoid problems with periodic boundary effects in z-direction, 
the thickness of the unit cell is chosen to be much larger during the simulation. 
The result is then corrected with respect to the true thickness. 
It is important to note here that the correction factor is the ratio
of the thickness of the simulation box relaxed in z-direction \rev{after equilibration} at the time of the start 
of strain and the actual thickness of the structure. 
\rev{Since the z-dimension also changes due to the barostat, 
we tested a dynamic correction factor and found that in the region we use here, 
neglecting this factor introduces only a negligible error compared to other error sources.}
The final result and its error are the averages of values in x- and y-direction 
and their corresponding errors. An engineering strain rate of $\varepsilon = 0.1$\,ps$^{-1}$ 
and metal units with a time step of $dt=0.0001$~ps are utilized for this simulations.
Simulations were carried out again at different temperatures
with and without pressure control in z-direction (\figref{graphene_baro}). 
Here a total strain of $0.02$, corresponding to 2000 simulated time steps, 
is considered as an elastic deformation and is therefore used to fit the linear function, 
see \figref{stressstrainbaro}.

First of all, it can be seen that the overall mechanical stability 
decreases slightly with temperature, since thermal fluctuations tend to soften a material. 
Overall, the values are compatible with literature values and 
results from the non-dynamic stress-strain method, compare Sec.~\ref{sec-3-2}. 

Apart from the first data point at the very low temperature of 1~K
almost all data points are well aligned and can be regarded as equal within the error.
The data point at 1~K probably captures a situation where classical 
molecular dynamics likely is no longer applicable.
When looking at the zoomed-in graph (\figref{graphene_baro} r.h.s.), 
slight deviations become visible, 
but these seem not to be systematic 
and therefore cannot be clearly attributed to the lack of pressure control. 
Upon examining the function of the Nos{\'{e}}-Hoover-barostat as implemented in LAMMPS, 
it becomes evident why no significant difference is observed. In the \textit{npt} ensemble, 
pressure in a given direction is regulated by adjusting the corresponding box dimensions \cite{LAMMPS:DOC}. 
However, since the structure is not directly coupled to the box size, this results in no appreciable effect.

\begin{figure}[ht!]
\centering
\includegraphics[width=0.49\textwidth]{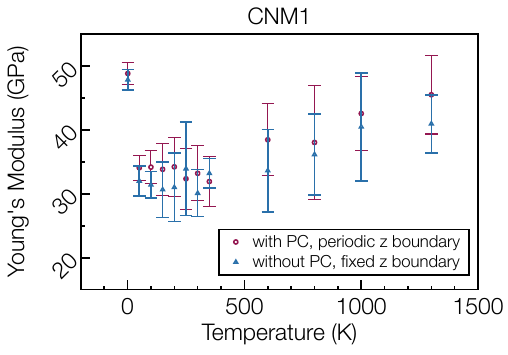}
\includegraphics[width=0.49\textwidth]{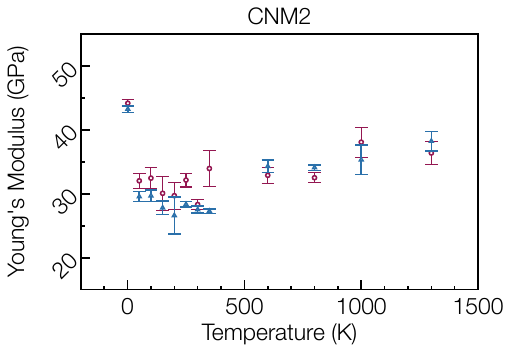}
\caption{Young's modulus depending on temperature, red circles correspond to a simulation 
with PBC and pressure control set to 1 bar in z-direction. 
The blue triangles represent data from simulation with fixed boundary 
and no pressure control in z direction. The error bars depict the error 
of the mean value of the x- and y-direction.\label{CNM_sumup}}
\end{figure}

Since carbon nanomembranes are quasi 2d structures with a non-vanishing 
thickness of about a nanometer \cite{YDB:AN18}, the next step is 
to investigate the influence of pressure control in z-direction. 
For this purpose, the Young's modulus of various CNMs was calculated; 
the CNMs were generated using the method described in \ref{sec-2-4}, see \figref{CNM_sumup}.

When CNMs are initialized with PBCs in z-direction, the same problem as with graphene occurs. 
Again the actual structure is not periodic in z-direction, while the box is. 
Pressure can therefore only be generated 
by a closer stacking of mirrored structures. However, this can turn problematic
as thermal fluctuation can lead to the formation of an artificial bulk phase
whose mechanical properties no longer correspond to those of the addressed CNM. 

In \figref{CNM_sumup} it can be seen that in most cases the value determined 
without barostat in z-direction, is lower. This would generally 
indicate that the stress is relieved via a different spatial direction 
and therefore does not only occur in the actively stretched direction. 
These disparities are, however, admittedly, relatively small compared 
to the large volume uncertainty. It is therefore not decisive whether a 
barostat is activated in z or not. 

Taking into account the errors, the applied temperature in a range of 50\,K to 350\,K 
does not seem to have any significant influence for the structures tested here. 
In general, this may of course be different.
The obtained mean values of the Young's modulus in this temperature regime are 
summarized in table \xref{baro-table-1} together with the values 
obtained by the other methods. 
\rev{An interesting side note is that, for CNMs, the Young's modulus, 
contrary to expectations, seems to increase with temperature. 
Perhaps this is a material property of CNMs. 
Since there is currently no research on this topic, 
it would be interesting to further investigate in this effect. 
What is known from literature, is that CNMs begin the process of 
graphitization at temperatures of the order of $10^3$~K \cite{AVW:ASCN13}. 
As temperature increases, some regions of the material could potentially 
transition towards more graphene-like structures, resulting in a higher tensile strength.}

\begin{table}[h!]
\centering
\caption{\label{baro-table-1}
Comparison of Young's moduli calculated with three distinct methods
for single-layer graphene (SLG) as well as carbon nanomembranes 1 \& 2, 
compare Figs.~\xref{graphene} and \xref{CNM}, for temperatures in the range of 50~K to 350~K.
All values are in GPa.} 
\begin{tabular}{|l|c|c|c|} %
\hline
\textbf{Method} & \textbf{SLG} & \textbf{CNM1} & \textbf{CNM2} \\
\hline
Scaling           &     $1088\pm11$        &    $240\pm5$         &     $253\pm3$  \\
\hline
Stress-Strain     &   $854\pm5.2$          &     $35.835\pm0.835$        & $35.71\pm5.13$     \\
\hline
Barostated dyn., PC in z &     $854\pm23$        &     $33\pm5$        &  $31\pm3$   \\
\hline
Barostated dyn., without PC in z&   $854\pm19$          &  $32\pm7$ & $28\pm3$            \\
\hline
\end{tabular}
\label{table}
\end{table}

\section{Summary}
\label{summary}

If more than a rough estimate of the Young's modulus of crystalline 2d or quasi 2d structures, 
or even the tensile strength of an amorphous structure is intended, the 
\rev{popular} scaling method 
with its calculation via the curvature of the potential energy should not be favoured. 
While the value for single-layer graphene is in agreement with calculations from reference \cite{GEH:PE18}, 
the moduli for 
\rev{disordered}
CNM structures are an order of magnitude too big compared to experimental 
values of about 10\,GPa \cite{ZBG:B11} as already reported by Ehrens et al. \cite{EGV:PRB21}. 
Both stress-strain based algorithms perform much better and more reliable 
as they also account for other elastic deformation mechanism instead of only bond elongation. 
The obtained results are much closer to the predicted 10\,GPa for CNMs.
Remaining discrepancies are probably due to the uncertainties of the structure itself.

Which \rev{of the investigated common} methods works best depends on the application. 
The clamped region stress-strain method 
is easier and saver to use, but only calculations at $T=0$ are possible.
If it is assumed that the mechanical stability varies noticeably with temperature, 
this method should not be used.

The barostated dynamics method on the other hand, 
allows for calculation under several temperature and pressure conditions, 
but is more difficult to adjust due to the larger number of parameters,
and it requires closer monitoring \cite{CDB:MS17}. 

Finally, considering the barostat method, if no change in pressure in z-direction is observable 
for (quasi) 2d materials during straining, 
even without activated pressure control,
we would recommend not to use pressure control for (quasi) 2d materials, 
as this could be a possible source of errors. 

\rev{In conclusion, with our investigation we provide step-by-step instructions 
how to perform trustworthy determinations of the Young's modulus.}

\section*{Acknowledgments}

We thank Nigel Marks and Fil Vukovic for fruitful discussions.


\end{document}